\documentclass[24pt]{article}
\usepackage{geometry}
\usepackage{xcolor}
\usepackage{a4}
\usepackage{alltt}
\usepackage{graphicx}
\usepackage{epsf}
\usepackage{amsmath}
\usepackage{amssymb}
\usepackage{cite}
\usepackage{hyperref}
\usepackage{blindtext}
\usepackage[toc,page]{appendix}

\newcommand{\be}{\begin{equation}}
\newcommand{\ee}{\end{equation}}

\newcommand{\Rmnum}[1]{\expandafter\@slowromancap\romannumeral #1@}
\newcommand{\bea}{\begin{eqnarray}}
\newcommand{\eea}{\end{eqnarray}}
\begin{document}

\title{Reconstruction method of $f(R)$ gravity for isotropic and anisotropic spacetimes}
\author{Saikat Chakrabarty
\thanks{E-mail:~ snilch@iitk.ac.in} \\ 
\normalsize Department of Physics, Indian Institute of Technology,\\ 
\normalsize Kanpur 208016, India}
\maketitle

\begin{abstract}
We present the reconstruction method of $f(R)$ gravity for the homogeneous and anisotropic Bianchi-I spacetime, which was previously formulated only for homogeneous and isotropic FLRW spacetime. We argue in this paper that for anisotropic spacetimes, the total anisotropy behaves as an independent metric degree of freedom on top of the average scale factor in $f(R)$ gravity. This is not like $GR$, where specifying the form of the average scale factor as a function of time also specify the total anisotropy as a function of time uniqely. We link this peculiar fact to an interesting intertwining between the definition of Ricci scalar for anisotropic metric and anisotropy evolution equation in $f(R)$ gravity. Consequently, specifying an anisotropic solution of $f(R)$ gravity implies specifying both the average scale factor and the total anisotropy as functions of time. The reconstruction method hence formulated is applied to two scenarios where anisotropy suppression is important, namely, a quasi de-Sitter expansion as required in inflation, and a power law contraction as required in ekpyrotic bounce models.
\end{abstract}

\section{Introduction}

The present universe is spatially homogeneous and isotropic in large scales to a high degree of accuracy as far as the observational evidences go, and can be very well approximated by the Friedmann-Lemaitre-Robertson-Walker ($FLRW$) metric. Still, possibility of some amount of spatial anisotropy in some early epochs of the universe cannot be ruled out. In fact, any 'good' physical theory must be general enough in nature so that the homogeneous and isotropic universe arises as an attractor solution in the phase space of more general solutions. Even if the initial spacetime is anisotropic, there must be some mechanism of anisotropy suppression in the theory one considers. We try to tackle the problem in the reverse way. Provided that an anisotropy suppression occurs, we present a formalism to reconstruct a 'good' theory of gravity. 

Usually any pre-existing anisotropy is assumed to be small and incorporated only as cosmological perturbations in the metric. Small metric anisotropy has traditionally been analyzed by the techniques of dynamical system analysis applied to cosmology (for example, see references \cite{Wainwright:1998ms, Leon:2014dea, Leon:2010pu, DeFelice:2013awa, Barrow:2000ka}). A realistic isotropic cosmology must be dynamically stable with respect to small anisotropic perturbations. In other words a realistic isotropic cosmology must arise as an attractor solution. This approach has found huge application in the field of inflationary cosmology [\cite{Chen:2001fh, Pereira:2015pga, Anninos:1991ma, Kitada:1992uf}]. A great example is the cosmic no hair conjecture, which states that the de-Sitter solution must be an attractor solution under general relativistic dynamics for all homogeneous anisotropic models. Wald has been able to prove it for all Bianchi models except Bianchi-IX (see \cite{Wald:1983ky}). Attractor nature of the de-Sitter solution has also been studied in Starobinski's inflationary scenario, which incorporates $R+\alpha R^2$ gravity \cite{Maeda:1987xf, Cotsakis:1993er, Schmidt:1988hq}. Another area where an isotropization mechanism becomes important is in the context of bouncing scenario \cite{Garfinkle:2008ei, Bozza:2009jx, Barrow:2010rx, Barrow:2015wfa}, which is an alternative to inflation. It is well known that in $GR$ spatial anisotropy evolves as $a^{-6}$. So in an expanding universe, anisotropy dies out quickly, but in a contracting universe, anisotropy actually grows large with time, which, if not suppressed, might foil a subsequent bounce. This is the case of well known $BKL$ instability \cite{Belinsky:1970ew}. To restore the situation in the context of $GR$, usually an ekpyrotic phase dominated by some ultrastiff matter component is invoked. An ultrastiff matter component has an equation of state greater than unity, so in a contracting universe the energy density of the ultrastiff matter grows faster than the anisotropy, and effectively suppresses it. Considering gravity theories beyond $GR$, evolution of anisotropy in a contracting universe has some interesting characteristics for $R+\alpha R^2$ gravity, which has recently been studied in  \cite{Bhattacharya:2017cbn}.

In this paper also we deal with metric anisotropy in $f(R)$ gravity, but we take a different approach. We assume a Bianchi-I model with some specific form of the average scale factor together with some anisotropy that is getting exponentially suppressed with respect to the average Hubble parameter. We then show that, given the equation of state parameter of the dominant matter contribution, it is possible to reconstruct the $f(R)$ gravity. Although getting a compact form of $f(R)$ may not be always be possible, nevertheless it is possible at least to reconstruct the function $f(R)$ term by term in a systematic manner. Reconstruction method of $f(R)$ gravity has been used previously in literature for realizing some specific cosmological evolutions in the isotropic $FLRW$ case \cite{Nojiri:2008nk, Nojiri:2009kx, Bamba:2013fha, Nojiri:2006gh, Nojiri:2006be, Nojiri:2017ncd}. 

The paper is organized as follows. In section \ref{2}, we briefly review the reconstruction method of $f(R)$ gravity for the isotropic case, followed by it's application to toy models of matter bounce and inflation in section \ref{3}. In section \ref{4}, we formulate the reconstruction method of $f(R)$ gravity in presence of metric anisotropy, followed by it's application to the case of a power law contraction (mimicking a pre-bounce ekpyrotic phase) and a quasi de-Sitter expansion (mimicking an inflation) in section \ref{5}. Next we conclude with some discussion of the behavior of anisotropy in $f(R)$ gravity and the and it's relevance to the present method.

\section{Reconstruction method for isotropic case}
\label{2}

Let us briefly review the reconstruction method of $f(R)$ gravity for the homogeneous and isotropic  FLRW metric 
\begin{equation}
ds^2=-dt^2+a^2(t)[dx_1^2+dx_2^2+dx_3^2],
\label{flrw}
\end{equation}
where symbols carry the usual meanings. The equations of it's dynamics under $f(R)$ gravity in presence of an isotropic fluid are
\begin{eqnarray}
3H^2&=&\frac{\kappa}{f^{\prime}}\left(\rho+\rho_{\rm{curv}}\right),
\label{flrw_constraint_1}\\
2\dot{H}+3H^2&=&-\frac{\kappa}{f^{\prime}}\left(p+p_{\rm{curv}}\right),
\label{flrw_dynamic_1}\\
p-\omega\rho &=&0,
\label{flrw_constraint_2}\\
\dot{\rho}+3H\left(\rho+p\right)&=&0,
\label{flrw_dynamic_2}
\end{eqnarray}
where in the above equations
\begin{eqnarray}
\rho_{\rm{curv}}&=&\frac{Rf^{\prime}-f}{2\kappa}-\frac{3Hf^{\prime\prime}\dot{R}}{\kappa},
\label{rho_curv}
\\
p_{\rm{curv}}&=&\frac{\dot{R}^2f^{\prime\prime\prime}+2H\dot{R}f^{\prime\prime}+\ddot{R}f^{\prime\prime}}{\kappa}-\frac{Rf^{\prime}-f}{2\kappa}.
\label{p_curv}
\end{eqnarray}
Here an overdot denotes time derivative and prime denotes derivative with respect to $R$. In the above, note that, once we know the form of the function $f(R)$, there are a total of three functions of time, $H(t),\,\rho(t),\,p(t)$ that governs the dynamics. Existence of the two constraint equations \ref{flrw_constraint_1} and \ref{flrw_constraint_2} implies that only one of these three functions is independent, while the other two can be determined using the constraint equations. For example, given a form of $f(R)$, we can use the constraint equations to write the dynamical equation \ref{flrw_dynamic_1} as a differential equation in only one single function $H(t)$. Once we solve for the function $H(t)$, the functions $\rho(t)$ and $p(t)$ can be found using the two constraint equations.
 
However, there may be cases when the form of $f(R)$ is not known a-priori. Given some value of the parameter $\omega$, we would like to derive a suitable form of $f(R)$ so as to realize a particular solution $a(t)$. This program, i.e., specifying the solution $a(t)$, the value of the parameter $\omega$ and finding the form $f(R)$ that can realize this as a solution is called the reconstruction method. For the isotropic case it works as follows. Using the definition of $\rho_{\rm{curv}}$ as in equation \ref{rho_curv}, the constraint equation \ref{flrw_constraint_1} can be written as a differential equation for $f(R)$,
\begin{equation}
3H\dot{R}f^{\prime\prime}(R)+\left(3H^2-\frac{R}{2}\right)f^{\prime}(R)+\frac{1}{2}f(R)-\kappa\rho_0 a^{-3(1+\omega)}=0.
\label{fr_recon_flrw}
\end{equation}
Let us define $N\equiv\ln a(t)$, where the scale factor $a(t)$ is a known function of time. This relation can be inverted to give $t=t(N)$. We have then the following,
\begin{eqnarray}
H(t)&=&\dot{N}(t),
\\
H^2(t(N))&=&\dot{N}^2(t)=\dot{N}^2(t(N)).
\end{eqnarray}
Let $H^2(t(N))=G(N)$. Therefore
\begin{eqnarray}
\dot{H}(t(N))&=&\frac{1}{2}G^{\prime}(N),
\\
R(t(N))&=&6\left(\dot{H}+2H^2\right)=3G^{\prime}(N)+12G(N),
\label{R(t(N))_flrw}\\
\dot{R}(t(N))&=&\frac{dR(t)}{dN}\dot{N}(t)=\left(3G^{\prime\prime}(N)+12G^{\prime}(N)\right)H(t),
\\
H\dot{R}&=&\left(3G^{\prime\prime}(N)+12G^{\prime}(N)\right)H^2(t)=3G(N)\left(G^{\prime\prime}(N)+4G^{\prime}(N)\right).
\end{eqnarray}
Equation \ref{R(t(N))_flrw} can be inverted to yield $N(R)$. Then the equation \ref{fr_recon_flrw} can be re-written as
\begin{equation}
9G(N(R))\left[G^{\prime\prime}(N(R))+4G^{\prime}(N(R))\right]f^{\prime\prime}(R)-3\left[G(N(R))+\frac{1}{2}G^{\prime}(N(R))\right]f^{\prime}(R)+\frac{1}{2}f(R)=\kappa\rho_0 e^{-3N(R)(1+\omega)}.
\label{flrw_recon}
\end{equation}
For vacuum situation the R.H.S. of the above equation vanish. So the equation becomes homogeneous and solving it becomes relatively easier.

\section{Examples for isotropic case}
\label{3}

In this section the reconstruction method of $f(R)$ gravity for isotropic case, as discussed in the previous section is applied in a bouncing scenario and in a quasi de-Sitter inflationary scenario.

\subsection{Exponential bounce model : $a(t)=e^{Ct^2}$($C>0$)}

Reconstruction of $f(R)$ gravity for this model has been done in \cite{Bamba:2013fha}. For this model,
\begin{eqnarray}
N&\equiv &\ln a=Ct^2,\\
H&=&\dot{N}=2Ct,\\
H^2&=&4CN\equiv G(N),\\
R(N)&=&12C(1+4N),\\
G(N(R))&=&\frac{R}{12}-C.
\end{eqnarray} 
The reconstruction equation becomes
\begin{equation}
144C(R-12C)\frac{d^2f}{dR^2}-3(R+12C)\frac{df}{dR}+6f=12\kappa\rho_0\exp\left[-\frac{3}{4}(1+\omega)\left(\frac{R}{12C}-1\right)\right].
\label{fr_recon_exp_bounce}
\end{equation}
If the matter contribution can be neglected, then the above equation admits two polynomial solutions as follows,
\begin{equation}
f(R)=-2C+R-\frac{1}{72C}R^2,\,\,\,(R-12C)^{3/2} L_{\frac{1}{2}}^{\frac{3}{2}}\left(\frac{R}{48 C}-\frac{1}{4}\right),
\end{equation}
where $L_n^a(x)$ is the generalized Laguerre polynomial. An $R^2$ theory with a negative coefficient of $R^2$, such as the first one, was shown to produce bouncing solutions in \cite{Paul:2014cxa}. A general solution would be a linear combination of the two solutions shown above.

The matter contribution usually can not be neglected in a matter bounce scenario. In that case equation \ref{fr_recon_exp_bounce} is inhomogeneous and finding the compact form of a particular solution becomes much more complicated. Nevertheless, an approximate compact form of $f(R)$ can be obtained in the limit $N\approx 0$ (near the bounce) and in the limit $N\rightarrow\infty$ (away from the bounce). Near the bounce $R\approx 12C$, so that equation \ref{fr_recon_exp_bounce} becomes
\begin{equation}
72f^{\prime}(R)-6f(R)=-12\kappa\rho_0,
\end{equation}
which has the solution,
\begin{equation}
f(R)=2\kappa\rho_0+C_1e^{R/12},
\end{equation}
for some constant of integration $C_1$. Away from the bounce $R\gg 12C$, so that Eq.\ref{fr_recon_exp_bounce} becomes
\begin{equation}
144CRf^{\prime\prime}(R)-3Rf^{\prime}(R)+6f(R)=0,
\end{equation}
which admits a polynomial solution as follows,  
\begin{equation}
f(R)=R-\frac{R^2}{96C},\,\,\,R(96C-R)\rm{Ei}\left(\frac{R}{48C}\right)-48Ce^{\frac{R}{48C}} (48C-R).
\end{equation}
In the above 'Ei(x)' denote the exponential integral function, which, for real nonvanishing values of $x$, is defined as
\begin{center}
$\rm{Ei}(x)=-\int_{-x}^{\infty}\frac{e^{-t}}{t}dt$.
\end{center}

\subsection{Quasi de-Sitter expansion : $a(t)=e^{H_0t-\frac{M^2}{12}t^2}$($H_0>0$)}

For this model,
\begin{eqnarray}
N&\equiv &\ln a=H_0t-\frac{M^2}{12}t^2,\\
H&=&\dot{N}=H_0-\frac{M^2}{6}t,\\
H^2&=&H_0^2-\frac{M^2}{3}N\equiv G(N),\\
R(N)&=&12H_0^2-M^2-4M^2N,\\
G(N(R))&=&\frac{R+M^2}{12}.
\end{eqnarray} 
Concentrating only a vacuum dominated quasi de-Sitter expansion, the reconstruction equation becomes
\begin{equation}
4M^2(R+M^2)\frac{d^2f}{dR^2}+(R-M^2)\frac{df}{dR}-2f=0.
\label{fr_recon_infl}
\end{equation}
The above homogeneous equation has two linearly independent solutions as follows,
\begin{equation}
f(R)=\frac{M^2}{6}+R+\frac{1}{6M^2}R^2,\,\,\,e^{-\frac{R}{4 M^2}} \left(\sqrt{\pi } e^{\frac{1}{4} \left(\frac{R}{M^2}+1\right)} \left(M^4+6 M^2 R+R^2\right)\rm{Erf}\left(\frac{\sqrt{M^2+R}}{2 M}\right)+2 M \sqrt{M^2+R} \left(3 M^2+R\right)\right).
\end{equation}
In the above 'Erf(x)' denote the error function defined as
\begin{center}
$\rm{Erf}(x)=\frac{2}{\sqrt{\pi}}\int_{0}^{x}e^{-t^2}dt$.
\end{center}
The first of above solutions correspond to the well known inflationary model by Starobinsky \cite{Starobinsky:1980te}.

\section{Reconstruction method in the anisotropic case}
\label{4}

Let us now formulate the reconstruction method of $f(R)$ gravity for the homogeneous and isotropic  Bianchi-I metric.
\begin{equation}
ds^2=-dt^2+a^2(t)[e^{2\beta_1(t)}dx_1^2+e^{2\beta_2(t)}dx_2^2+e^{2\beta_3(t)}dx_3^2],
\label{bianchi-I}
\end{equation}
where $a(t)$ is the average scale factor and three functions $\dot{\beta}_1,\,\dot{\beta}_2,\,\dot{\beta}_3$ characterizes the deviation from the average scale factor in the three perpendicular dimensions. The total amount of anisotropy in the metric is given by the quantity $\dot{\beta}_1^2+\dot{\beta}_2^2+\dot{\beta}_3^2$. Observe that when $\dot{\beta}_1^2+\dot{\beta}_2^2+\dot{\beta}_3^2=0$, i.e., $\dot{\beta}_1^2=\dot{\beta}_2^2=\dot{\beta}_3^2=0$, the spatial coordinates can be suitably rescaled to recast the above metric in the FLRW form. The dynamics of the above metric for $f(R)$ gravity in presence of an isotropic fluid can be described by the following set of equations,
\begin{eqnarray}
3H^2&=&\frac{\kappa}{f^{\prime}}\left(\rho+\rho_{\rm{curv}}\right)+\frac{1}{2}\left(\dot{\beta_1}^2+\dot{\beta_2}^2+\dot{\beta_3}^2\right),
\label{bianchi_constraint_1}\\
2\dot{H}+3H^2&=&-\frac{\kappa}{f^{\prime}}\left(p+p_{\rm{curv}}\right)-\frac{1}{2}\left(\dot{\beta_1}^2+\dot{\beta_2}^2+\dot{\beta_3}^2\right),
\label{bianchi_dynamic_1}\\
p-\omega\rho &=&0,
\label{bianchi_constraint_2}\\
\dot{\rho}+3H\left(\rho+p\right)&=&0,
\label{bianchi_dynamic_2}\\
\ddot{\beta_1}+\left(3H+\frac{\dot{R}f^{\prime\prime}}{f^{\prime}}\right)\dot{\beta_1}&=&0,
\label{bianchi_dynamic_3}\\
\ddot{\beta_2}+\left(3H+\frac{\dot{R}f^{\prime\prime}}{f^{\prime}}\right)\dot{\beta_2}&=&0,
\label{bianchi_dynamic_4}\\
\ddot{\beta_3}+\left(3H+\frac{\dot{R}f^{\prime\prime}}{f^{\prime}}\right)\dot{\beta_3}&=&0,
\label{bianchi_dynamic_5}\\
\dot{\beta_1}+\dot{\beta_2}+\dot{\beta_3}&=&0,
\label{bianchi_constraint_3}
\end{eqnarray}
where $\rho_{\rm{curv}}$ and $p_{\rm{curv}}$ have the same definition as in the $FLRW$ case. Note that, once we know the form of the function $f(R)$, there are a total of six functions of time $H(t),\,\rho(t),\,p(t),\,\dot{\beta_1}(t),\,\dot{\beta_2}(t),\,\dot{\beta_3}(t)$ governing the dynamics. Existence of the three constraint equations \ref{bianchi_constraint_1}, \ref{bianchi_constraint_2}, \ref{bianchi_constraint_3} implies that only three of them are independent. From the constraint equation \ref{bianchi_constraint_3}, it can be concluded that only two of the $\dot{\beta}$'s are independent. However, observe that both in the constraint equation \ref{bianchi_constraint_1} and the dynamical equation \ref{bianchi_dynamic_1}, $\dot{\beta}$'s appear only as the combination $\dot{\beta_1}^2+\dot{\beta_2}^2+\dot{\beta_3}^2$, which is the total amount of anisotropy. Consequently, we conclude that it is only the total amount of anisotropy which enters into the dynamics. It is easy to check that the quantity $x$ defined by $x^2=\dot{\beta}_1^2+\dot{\beta}_2^2+\dot{\beta}_3^2$ obeys the a dynamical equation of the same form as obeyed by the $\dot{\beta}_i$'s themselves.
\begin{equation}
\dot{x}+\left(3H+\frac{\dot{R}f^{\prime\prime}}{f^{\prime}}\right)x=0.
\label{bianchi_dynamic_345}
\end{equation}
The reader is advised to be careful not to confuse the quantity $x(t)$ defined above with the notation $x$ used to denote the spatial coordinates of the metric as in equations \ref{flrw} or \ref{bianchi-I}.

If we know the form of the function $f(R)$ and concentrate only on the quantity $x(t)$, we see that there are now a total of four functions of time $H(t),\,\rho(t),\,p(t),\,x(t)$ governing the dynamics. Existence of the constraint equations \ref{bianchi_constraint_2} and \ref{bianchi_constraint_1} implies that only two of them are independent. Without loss of generality, we can choose them to be $H(t)$ and $x(t)$. Given some form of the function $f(R)$, they can be determined by solving equations \ref{bianchi_dynamic_1} and \ref{bianchi_dynamic_345}. $\rho(t)$ and $p(t)$ can then be found using the constraint equations \ref{bianchi_constraint_1} and \ref{bianchi_constraint_2}. However, if the form of $f(R)$ is not known a-priori, but a suitable form has to be determined so as to realize a particular solution, given a particular value of the parameter $\omega$, then we have to formulate the reconstruction program for the Bianchi-I metric in equation \ref{bianchi-I}. In this case, however, since $H(t),\,x(t)$ are two independent metric degrees of freedom, specifying a particular solution implies specifying both these functions. Usually, during a particular era given by the background evolution $a(t)$, we expect the dynamics to contain some kind of isotropization mechanism so as to match with our observed isotropic universe. In that case we can take $x(t)$ as,
\begin{equation}
\frac{x^2(t(N))}{H^2(t(N))}=e^{\lambda N},
\end{equation}
where $N\equiv\ln a$ as in the isotropic case, and the constant $\lambda$ is positive in a contracting universe and negative in an expanding universe (Assuming $a=1$ at time $t=0$). Formulation of reconstruction method in the anisotropic case now follows the same route as in the isotropic case. Equation \ref{bianchi_constraint_1} can be recast as a second order differential equation in $f(R)$,
\begin{equation}
3H\dot{R}f^{\prime\prime}(R)+\left(3H^2-\frac{R}{2}-\frac{1}{2}x^2\right)f^{\prime}(R)+\frac{1}{2}f(R)-\kappa\rho_0 a^{-3(1+\omega)}=0.
\label{fr_recon_bianchi}
\end{equation}
If $H^2(t(N))=G(N)$, then
\begin{eqnarray}
\dot{H}(t(N))&=&\frac{1}{2}G^{\prime}(N),
\\
R(t)&=&6\left(\dot{H}+2H^2\right)+x^2=3G^{\prime}(N)+G(N)(12+e^{\lambda N}),
\label{R(t(N))_bianchi}\\
\dot{R}(t)&=&\frac{dR(t)}{dN}\dot{N}(t)=[3G^{\prime\prime}(N)+G^{\prime}(N)(12+e^{\lambda N})+\lambda e^{\lambda N}G(N)]H(t),
\\
H\dot{R}&=&[3G^{\prime\prime}(N)+G^{\prime}(N)(12+e^{\lambda N})+\lambda e^{\lambda N}G(N)]H^2(t)\nonumber\\
&=&G(N)[3G^{\prime\prime}(N)+G^{\prime}(N)(12+e^{\lambda N})+\lambda e^{\lambda N}G(N)].
\end{eqnarray}
Equation \ref{R(t(N))_bianchi} can be inverted to yield $N(R)$. Then the equation \ref{fr_recon_bianchi} can be re-written as,
\begin{eqnarray}
&&3G(N(R))[3G^{\prime\prime}(N(R))+G^{\prime}(N(R))(12+e^{\lambda N(R)})+\lambda e^{\lambda N(R)}G(N(R))]f^{\prime\prime}(R)\nonumber\\
&&\,\,\,\,\,\,\,\,\,\,\,\,\,\,\,\,\,\,\,\,\,\,\,\,-\left[\left(3+\frac{1}{2}e^{\lambda N(R)}\right)G(N(R))+\frac{3}{2}G^{\prime}(N(R))\right]f^{\prime}(R)+\frac{1}{2}f(R)=\kappa\rho_0 e^{-3N(R)(1+\omega)}
\label{bianchi_recon}
\end{eqnarray}

\section{Examples for anisotropic case}
\label{5}

In this section the reconstruction method of $f(R)$ gravity for anisotropic case as discussed in the previous section is applied in isotropization scenario during an ekpyrotic power law contraction era and a quasi de-Sitter expansion era.

\subsection{Ekpyrotic contraction phase : $a(t)=(-t)^m$($t<0$, $0<m<1$)}

For this model,

\begin{eqnarray}
N&=&\ln a=m\ln(-t)\\
H(t)&=&\dot{N}=\frac{m}{t}\\
H^2(t(N))&=&\frac{m^2}{t^2}=\frac{m^2}{e^{2N/m}}\equiv G(N)\\
x^2(N)&=&G(N)e^{\lambda N}=m^2e^{(\lambda-2/m)N}\\
R(t(N))&=&6(\dot{H}+2H^2)+x^2=e^{-2N/m}[6m(2m-1)+m^2e^{\lambda N}]
\label{R(t(N))_power_law}
\end{eqnarray}

Equation \ref{R(t(N))_power_law} is in general not invertible except for the special cases $m=\frac{1}{2},\,\frac{2}{\lambda}$. If one chooses to concentrate on a sufficiently past epoch (i.e., $N$ is a very large positive quantity), then second term in $R(N)$ dominates, as $\lambda>0$. Essentially this means that in far past the leading contributor to the Ricci scalar is metric anisotropy. In this case equation \ref{R(t(N))_power_law} becomes approximately invertible and reconstruction program can be carried out. Note that, along the same line of reasoning, equation \ref{R(t(N))_power_law} is also invertible if one chooses to concentrate on an epoch close to the end of the power law contraction phase (i.e., $N$ is a very large negative quantity). In that case the contribution from the first term in $R(N)$ dominates. But this is essentially the situation when anisotropy becomes negligible w.r.t. the Hubble parameter, and consequently the problem becomes essentially reconstructing the $f(R)$ gravity for an isotropic evolution.

Concentrating on a sufficiently past epoch when the anisotropy is much larger than the Hubble parameter, equation \ref{R(t(N))_power_law} approximately becomes
\begin{equation}
R(N)\approx m^2e^{\left(\lambda -2/m\right)N},
\end{equation}
which can be inverted to obtain
\begin{equation}
N(R)=\left(\frac{m}{m\lambda-2}\right)\ln\left(\frac{R}{m^2}\right).
\end{equation}
Therefore, 
\begin{equation}
G(N(R))=m^2e^{-2N/m}=m^{2m\lambda/(m\lambda-2)}R^{-2/(m\lambda-2)}.
\end{equation}
Let us define $\gamma=\frac{2}{m\lambda-2},\,k=2-\frac{3}{2}m(1+\omega)$. After some pretty straightforward algebra, the reconstruction equation \ref{bianchi_recon} can be recast in the following form, 
\begin{equation}
A(R)f^{\prime\prime}(R)+B(R)R^{\gamma}f^{\prime}(R)+R^{2\gamma}f(R)=C R^{k\gamma},
\label{fr_recon_ekpyrotic}
\end{equation}
where the coefficients $A(R),\,B(R)$ has the form
\begin{equation}
A(R)=a_0+a_1R^{1+\gamma},\,\,\,\,B(R)=b_0+b_1R^{1+\gamma},
\end{equation}
and $a_0,\,a_1,\,b_0,\,b_1$ and $C$ are parameter dependent constants having the following values
\begin{eqnarray}
&&a_0=72m^{3+4\gamma},\,\,\,a_1=\frac{12}{\gamma}m^{2+2\gamma},\nonumber\\
&&b_0=6m^{1+2\gamma},\,\,\,b_1=-1,\nonumber\\
&&C=2\kappa\rho_0 m^{3\gamma m(1+\omega)}.
\end{eqnarray}
To check whether a power law solution in $R$ exists, one must check the nature of the point $R=0$ for Eq.\ref{fr_recon_ekpyrotic}. Let us define
\begin{equation}
P(R)=R^{\gamma}\frac{B(R)}{A(R)}=R^{\gamma}\frac{b_0+b_1R^{1+\gamma}}{a_0+a_1R^{1+\gamma}},\,\,\,\,\,\,Q(R)=R^{2\gamma}\frac{1}{A(R)}=\frac{R^{2\gamma}}{a_0+a_1R^{1+\gamma}},
\end{equation}
and then note that, as $R\rightarrow 0$,
\begin{itemize}
\item For $\gamma>0,\,\,\,P(R)\rightarrow 0,\,Q(R)\rightarrow 0$ ; $R=0$ is an ordinary point.
\item For $-1<\gamma<0,\,\,\,RP(R)\rightarrow 0,\,R^2Q(R)\rightarrow 0$ ; $R=0$ is a regular singular point.
\item For $\gamma=-1$, $RP(R)$ and $R^2Q(R)$ tends to some finite quantity ; $R=0$ is a regular singular point.
\item For $\gamma<-1,\,\,\,RP(R)\rightarrow\infty$ ; $R=0$ is an irregular singular point.
\end{itemize}
Therefore, in principle, for a power series solution to Eq.\ref{fr_recon_ekpyrotic} about $R=0$ to exist, $\gamma$ must satisfy $\gamma\geq -1$. Since the parameters $m$ and $\lambda$ are both positive in this case, from the definition of $\gamma$, this implies $m$ and $\lambda$ has to be chosen such that $\lambda>\frac{2}{m}$. Recalling that $x^2(N)=m^2e^{(\lambda-2/m)N}$, this implies that anisotropy is not only getting suppressed with respect to the Hubble parameter, but actually diminishing with time. Since $\gamma,\,k$ and $k\gamma$ not necessarily have to be integers, it is difficult to find a generic series solution about $R=0$ or a polynomial solution. In the special case when the rate of anisotropy suppression is much greater than the rate of Hubble parameter increment during contraction ($\lambda\gg\frac{2}{m}$), then $\gamma\rightarrow 0$ and $R\gg 1$. In this case,
\begin{equation}
A(R)\approx 6m^3\lambda R,\,\,\,\,B(R)\approx -R,\,\,\,\,C=2\kappa\rho_0,
\end{equation}
and the general solutions of Eq.\ref{fr_recon_ekpyrotic} is,
\begin{equation}
f(R)=2\kappa\rho_0+C_1R+C_2R\left(\frac{\rm{Ei}\left(\frac{R}{6 m^3 \lambda }\right)}{6 \lambda  m^3}-\frac{e^{\frac{R}{6 \lambda  m^3}}}{R}\right),
\end{equation}
where $C_1$ and $C_2$ are two integration constants. In the above 'Ei(x)' denotes exponential integral function as defined previously.

The cases $m=\frac{2}{\lambda},\,\frac{1}{2}$ are described separately in the two following subsections.

\subsubsection{Case I : $\lambda=\frac{2}{m}$($m\neq\frac{1}{2}$)}

In this case, from Eq.\ref{R(t(N))_power_law}, 
\begin{equation}
R(t(N))=e^{-2N/m}[6m(2m-1)+m^2e^{2N/m}]=6m(2m-1)e^{-2N/m}+m^2.
\end{equation}
Inverting this, 
\begin{equation}
N(R)=\frac{m}{2}\ln\left[\frac{6m(2m-1)}{R-m^2}\right],
\end{equation}
and therefore
\begin{equation}
G(N(R))=m^2e^{-2N/m}=\frac{m(R-m^2)}{6(2m-1)}.
\end{equation}
The reconstruction equation \ref{bianchi_recon} becomes of the form
\begin{equation}
2(R-m)^2f^{\prime\prime}(R)+(2m-1)(R-2m^3)f^{\prime}(R)+(2m-1)^2f(R)=C(R-m^2)^{2-k},
\label{fr_recon_ekpyrotic_I}
\end{equation}
with 
\begin{equation}
C=\frac{\kappa\rho_0}{18m^2}[6m(2m-1)]^k,\,\,\,k=2-\frac{3}{2}m(1+\omega).
\end{equation}
$R=0$ is an ordinary point of the differential equation \ref{fr_recon_ekpyrotic_I}, meaning a series solution about $R=0$ is, although very complicated, still in principle obtainable. The general solution to the homogeneous part of the equation comes in terms of Kummer confluent hypergeometric functions. In the far past epoch ($N\gg 0$), however, $R\rightarrow m^2$, and in this limit Eq.\ref{fr_recon_ekpyrotic_I} becomes approximately homogeneous. Then the two linearly independent solutions become
\begin{equation}
f(R)\sim\exp\pm\left[\frac{(2 m-1) \left(2 m^2+\sqrt{4 m^4-4 m^3-7 m^2+16 m-8}-m\right) R}{4 (m-1)^2 m}\right].
\end{equation}

\subsubsection{Case II : $m=\frac{1}{2}$ }

In this case, from equation \ref{R(t(N))_power_law}, we get, 
\begin{equation}
R(t(N))=\frac{1}{4}e^{(\lambda-4)N}.
\end{equation}
Inverting this, 
\begin{equation}
N(R)=\frac{1}{\lambda-4}\ln(4R),
\end{equation}
and therefore,
\begin{equation}
G(N(R))=\frac{1}{4}e^{-4N}=4^{\lambda/(4-\lambda)}R^{4/(4-\lambda)}.
\end{equation}
The reconstruction equation \ref{bianchi_recon} becomes of the form
\begin{equation}
A(R)f''(R)-B(R)f'(R)+f(R)=CR^{3(1+\omega)/(4-\lambda)},
\label{fr_recon_ekpyrotic_II}
\end{equation}
where the coefficients $A(R)$, $B(R)$ and the constant $C$ are,
\begin{eqnarray}
A(R)&=&6\times 4^{\lambda/(4-\lambda)}R^{4/(4-\lambda)}\left[12\times 4^{4/(4-\lambda)}R^{4/(4-\lambda)}+(\lambda-4)R\right],\nonumber\\
B(R)&=&R-3\times 4^{4/(4-\lambda)}R^{4/(4-\lambda)},\nonumber\\
C&=&2\kappa\rho_0 4^{3(1+\omega)/(4-\lambda)}.
\end{eqnarray}
To check whether a power law solution in $R$ exists, one must check the nature of the point $R=0$ for Eq.\ref{fr_recon_ekpyrotic_II}. Let us define
\begin{equation}
P(R)=\frac{B(R)}{A(R)},\,\,\,\,\,\,Q(R)=\frac{1}{A(R)},
\end{equation}
and recall that $\lambda>0$. It is straightforward to check that, as $R\rightarrow 0$,
\begin{itemize}
\item For $\lambda>4,\,\,\,RP(R)\rightarrow 0,\,R^2Q(R)\rightarrow 0$ ; $R=0$ is a regular singular point.
\item For $0<\lambda<4,\,\,\,RP(R)\rightarrow\infty,\,R^2Q(R)\rightarrow\infty$ ; $R=0$ is a irregular singular point.
\end{itemize}
Therefore for a power series solution in $R$ to exist, $\lambda$ must satisfy $\lambda>4$. Again, finding a generic compact form of $f(R)$ may not be possible. In the special case when the rate of anisotropy suppression is much greater than the rate of Hubble parameter increment during contraction ($\lambda\gg 4$), then $R\gg 1$. In this case,
\begin{equation}
A(R)\approx \frac{3}{2}\lambda R,\,\,\,\,B(R)=R,\,\,\,\,C=2\kappa\rho_0,
\end{equation}
and the general solutions is,
\begin{equation}
f(R)=\kappa\rho_0+C_1R+C_2R\left(\frac{2\rm{Ei}\left(\frac{2 R}{3\lambda }\right)}{3\lambda }-\frac{e^{\frac{2R}{3\lambda }}}{R}\right),
\end{equation}
where $C_1$ and $C_2$ are two integration constants. In the above 'Ei(x)' denotes exponential integral function as defined previously.

\subsection{Quasi de-Sitter expansion : $a(t)=e^{H_0t-\frac{M^2}{12}t^2}$($H_0>0$)}

For this model, $N$ and $G(N)$ bears the same definition as in the isotropic case. However, in the anisotropic case,
\begin{equation}
R(N)=-M^2+\left(H_0^2-\frac{1}{3}M^2N\right)(12+e^{\lambda N}),
\label{R(t(N))_infl}
\end{equation}
where $\lambda$ is negative in this case. Note that the above expression is not in general invertible. For small $N$, i.e., within a very small time from the beginning of the expansion, equation \ref{R(t(N))_infl} is approximately invertible and we obtain,
\begin{equation}
N(R)=3\left(\frac{R+M^2-13H_0^2}{3\lambda H_0^2-13M^2}\right),
\end{equation}
and therefore,
\begin{equation}
G(N(R))=\frac{M^2R+M^4-3\lambda H_0^4}{13M^2-3\lambda H_0^2}.
\end{equation}
Concentrating only a vacuum dominated quasi de-Sitter expansion, the reconstruction equation takes the form,
\begin{equation}
[c_1(R+M^2)+c_2]\frac{d^2f}{dR^2}-[c_3(R+M^2)+c_4]\frac{df}{dR}+f=0,
\label{fr_recon_infl_anisotropic}
\end{equation}
where $c_1$, $c_2$, $c_3$, $c_4$ are model dependent constants as follows,
\begin{eqnarray}
c_1=2\left(\frac{9\lambda^2 H_0^4+13M^4-9\lambda H_0^2M^2}{3\lambda H_0^2-13M^2}\right),\,\,\,c_2=-6\lambda H_0^4\left(\frac{36\lambda H_0^2-13M^2}{3\lambda H_0^2-13M^2}\right),\nonumber\\
c_3=\frac{3\lambda H_0^2-7M^2}{3\lambda H_0^2-13M^2},\,\,\,c_4=-\frac{18\lambda H_0^4-13M^4+3\lambda H_0^2M^2}{3\lambda H_0^2-13M^2}.
\end{eqnarray}
The Hubble slow roll parameter $\epsilon$ is defined as $\epsilon=-\frac{\dot{H}}{H^2}\approx\frac{M^2}{6H_0^2}$. As shown in reference \cite{DeFelice:2010aj}, solving the horizon and flatness problem of the big bang cosmology requires the inflation to occur for at least $70$ e-foldings, which places the lower upper bound on the Hubble slow roll parameter as $\epsilon\lesssim 7\times 10^{-3}$. Taking $\epsilon=7\times 10^{-3}$ gives $H_0=5M$. Therefore,
\begin{eqnarray}
c_1=2M^2\left(\frac{5625\lambda^2 -225\lambda +13}{75\lambda -13}\right),\,\,\,c_2=-3750\lambda M^4\left(\frac{900\lambda -13}{75\lambda -13}\right),\nonumber\\
c_3=\frac{75\lambda -7}{75\lambda -13},\,\,\,c_4=-M^2\left(\frac{11325\lambda -13}{75\lambda -13}\right).
\end{eqnarray}

Close to the start of the quasi de-Sitter epoch ($N\approx 0$), $R\approx 13H_0^2-M^2\approx 324M^2$. In this limit, equation \ref{fr_recon_infl_anisotropic} admits two linearly independent solutions which are as follows,
\begin{equation}
f(R)\sim\exp \left[\frac{R \left(\pm\sqrt{8219 M^4-3750 \lambda  M^4}+87 M^2\right)}{50 \left(75 \lambda  M^4-13 M^4\right)}\right].
\end{equation}

\section{A remark on the reconstruction method}

In this section we make a general remark on the reconstruction method we have used. Observe that, both for isotropic or anisotropic case, the reconstruction equation is a second order ordinary differential equation. Therefore the general solution to a reconstruction equation is of the form
\begin{equation}
f(R)=f_0(R)+C_1f_1(R)+C_2f_2(R),
\end{equation} 
where $f_0(R)$ is the particular integral, $f_1(R)$ and $f_2(R)$ are complimentary functions of the differential equation and $C_1$, $C_2$ are arbitrary integration constants. The particular integral appears because of the matter contribution. When matter contribution can be neglected, the reconstruction equation is homogeneous and the general solution can be expressed as the linear combination of the complimentary functions. Let us assume that the functions $f_0(R)$, $f_1(R)$, $f_2(R)$ can be Taylor expanded around $R=0$ as
\begin{center}
$f_0(R)=f_0(0)+f_0^{\prime}(0)R+\frac{1}{2}f_0^{\prime\prime}(0)R^2+...$,\\
$f_1(R)=f_1(0)+f_1^{\prime}(0)R+\frac{1}{2}f_1^{\prime\prime}(0)R^2+...$,\\
$f_0(R)=f_2(0)+f_2^{\prime}(0)R+\frac{1}{2}f_2^{\prime\prime}(0)R^2+...$.
\end{center}
Therefore 
\begin{center}
$f(R)=(f_0(0)+C_1f_1(0)+C_2f_2(0))+(f_0^{\prime}(0)+C_1f_1^{\prime}(0)+C_2f_2^{\prime}(0))R+\frac{1}{2}(f_0^{\prime\prime}(0)+C_1f_1^{\prime\prime}(0)+C_2f_2^{\prime\prime}(0))...$.
\end{center}
The choice of the two integration constants $C_1$ and $C_2$ is at our hand. They can always be chosen such as to get any desired value of the cosmological constant, and to scale the coefficients such that the coefficient of $R$ is unity. Suppose we want the $f(R)$ obtained by solving the reconstruction equation to reduce to $-2\Lambda +R$ in low curvature limit(where $\Lambda>0$ is the cosmological constant). Then we must set
\begin{eqnarray}
f_0(0)+C_1f_1(0)+C_2f_2(0)&=&-2\Lambda,\\
f_0^{\prime}(0)+C_1f_1^{\prime}(0)+C_2f_2^{\prime}(0)&=&1.
\end{eqnarray}
These two equations can then be solved to get the desired numerical values of the integration constant $C_1$ and $C_2$, so that our theory successfully reduces to $-2\Lambda +R$ in low curvature limit. 

Also observe that the above pair of linear equations in $C_1$, $C_2$ are similar to what we would have obtained by demanding $f(0)=-2\Lambda$ and $f'(0)=1$. Therefore, when solving a reconstruction differential equation numerically, it may not even be required to find the particular integral and the complimentary functions separately. All we need is to solve the differential equation numerically by imposing the initial conditions
\begin{center}
$f(0)=-2\Lambda\,\,,\,\,f'(0)=1$.
\end{center}

\section{Discussion and conclusion}

In the present work we try to extend the reconstruction technique of $f(R)$ gravity, which was previously formulated for homogeneous and isotropic FLRW background, to the more general case of homogeneous and anisotropic Bianchi-I background. Given some specific anisotropic background solution, it is possible to reconstruct the functional form of the action $f(R)$. We again emphasize that in general it might be hard to find a compact functional form in most of the cases, but the modifications to the Einstein-Hilbert action can at least be found term by term. It must be emphasized that in a given problem just plain application of the reconstruction method to find out a solution for $f(R)$ is not enough. One has to check for various conditions for the physical viability of the $f(R)$ solution obtained, viz. $f^{\prime}(R)>0$, $f^{\prime\prime}(R)>0$. 

Moreover, stability of the solution is another issue that needs to be dealt with. Once the $f(R)$ gravity corresponding to an anisotropic solution is obtained, the stability of the solution under that theory of gravity can be checked using the same method as for the isotropic case (e.g.see \cite{Bamba:2013fha}, \cite{Nojiri:2010wj}). For the isotropic case the stability analysis is done by examining the behavior of a small perturbation $\delta G(N)$ to the background solution $H^2(N)\equiv G(N)$. We can use the Friedmann equation to check whether the the perturbed solution goes back towards the background solution asymptotically. In the anisotropic case the stability analysis will be more involved. It is to be kept in mind that for the anisotropic case both $H^2(N)\equiv G(N)$ and $x^2(N)$ acts as independent metric degrees of freedom in $f(R)$ gravity. The form $x^2(N)=e^{\lambda N}H^2(N)$ was taken only on the assumption suppression of anisotropy with respect to the average Hubble parameter. A generic perturbation to the background solution will involve both $\delta(H^2(N))\equiv\delta G(N)$ and $\delta(x^2(N))$. A generic method of stability analysis of the anisotropic solutions needs to be formulated, which will be dealt with in a later publication.

An important point to note here is the behavior of metric anisotropy as an independent dynamic degree of freedom of the metric. This is in stark contrast with what happens in GR. In GR once the average scale factor $a(t)$ is given, the total amount of anisotropy $x(t)$ in a Bianchi-I metric is uniquely determined, since we have $x(t)\sim\frac{1}{a^3(t)}$. In $f(R)$ gravity, however, even if the form of $f(R)$ is known, the total amount of anisotropy $x(t)$ in a Bianchi-I metric is not uniquely determined in terms of the average scale factor $a(t)$. This is beause in $f(R)$ gravity we have $x(t)\sim\frac{1}{a^3(t)f^{\prime}(R)}$, so that the definition of $x$ contain $R$. But the definition of $R$ for an anisotropic metric itself contain $x$ in it, namely $R=6(\dot{H}+H^2)+x^2$. In fact, the definitions of $R$ and $x$ for $f(R)$ gravity are intertwined in such a way that, even if the functions $a(t)$ and $f(R)$ are known, it is impossible to find unique $R(t)$ or $x(t)$. Interested reader is referred to  \cite{Bhattacharya:2017cbn} for novelties of such anisotropic dynamics in metric $f(R)$ gravity, where it was shown that, even for the very simple case of $R+\alpha R^2$ gravity, there are in general three solutions for total anisotropy $x(t)$ possible for a given average scale factor $a(t)$. If the function $f(R)$ is not known a-priori, then the freedom of choosing $f(R)$ can be traded to obtain any $x(t)$ of our interest. Therefore, specifying a solution for the anisotropic case implies specifying both the functions $a(t)$ and $x(t)$.

Another important point, which we have emphasized in the last section, is that since the reconstruction equation is second order for both isotropic and anisotropic case, it yields two linearly independent solutions to the homogeneous part of the equation. Any linear combination of them is also a solution of the homogeneous part of the equation. It is always possible to choose the integration constants in a way to get any desired value of the cosmological constant and make the coefficient of $R$ is unity.   

In the present work, we have presented the formulation of the reconstruction method of $f(R)$ gravity assuming a particular time law for the scale factor. A shortcoming of this method is that even very simple time law of the scale factor may lead to differential reconstruction equation that has very complicated $f(R)$ solutions. It might even be impossible to find a general solution, as is the case with many of the examples we have considered. A different formulation of the reconstruction method of $f(R)$ gravity exists in literature, which is based not on any particular time law of the scale factor, but on actual expansion parameters that can be measured from the observations[\cite{Carloni:2010ph, Capozziello:2008qc, Capozziello:2014zda, Aviles:2012ir}]. This parameters are called cosmographic parameters. They are the Hubble rate, deceleration parameter, jerk, snap and lerk parameter, which are related to the first, second, third, fourth and fifth derivative of the scale factor respectively. In terms of this measured parameters the present time law of the scale factor can be found by doing a Taylor expansion around the present value of the scale factor, which is assumed to be unity. In an $f(R)$ cosmology the coefficients of the higher powers of $R$ in a series expansion of $f(R)$ can be related to these parameters. Therefore the observed values of these parameters also give bounds on the coefficients of the higher powers of $R$, and a series solution for $f(R)$ can be obtained. However, this approach to reconstruct $f(R)$ gravity has been formulated only for the isotropic case, as all the cosmographic parameters mentioned above are related to $FLRW$ cosmology. Extension of this approach to anisotropic Bianchi-I cosmology is still not found in literature.

\section{Acknowledgement} 

The author is indebted to Sergei Odintsov and Kazuharu Bamba for some valuable discussions, and to Arindam Mazumdar, Niladri Paul, Srimanta Bannerjee and Kaushik Bhattacharya for reviewing the draft.

%%%%%%%%%%%%%%%%%%%%%%%%%%%%%%%%%%%%%%%%%%%%%%%%%%%%%%%%%%%%%%%%%%%%%%%%%%%%%%%%%%%%%%%%%%%%%%%%%

\end{document}